\documentclass[preprint,aps]{revtex4}
\usepackage{graphicx}
\newcommand{\be}{\begin{equation}}
\newcommand{\ee}{\end{equation}}
\newcommand{\ba}{\begin{eqnarray}}
\newcommand{\ea}{\end{eqnarray}}
\newcommand{\p}{\partial}
\newcommand{\f}{\frac}

\begin{document}
\title
{Clarifying nonstatic-quantum-wave behavior through extending its analysis to the p-quadrature space:
Interrelation between the q- and p-space wave-nonstaticities
 \vspace{0.3cm}}
\author{Jeong Ryeol Choi\footnote{E-mail: choiardor@hanmail.net } \vspace{0.3cm}}

\affiliation{Department of Nanoengineering, Kyonggi University, 
Yeongtong-gu, Suwon,
Gyeonggi-do 16227, Republic of Korea \vspace{0.7cm}}

\begin{abstract}
If electromagnetic parameters of a medium vary in time,
quantum light waves traveling in it become nonstatic. A recent report shows
that such nonstatic waves can also appear even when the environment is static
where the parameters of the medium do not vary.
In this work, the properties of nonstatic waves
in a static environment are investigated
from their $p$-space analysis, focusing on the interrelation
between the $q$- and $p$-space nonstatic waves.
The probability densities in $p$-space (as well as in $q$-space) for both the
nostatic Fock and Gaussian states evolve in a periodic manner, i.e.,
they constitute belly and node in turn successively as time goes by.
If we neglect the displacement of waves,
the $q$- and $p$-space wave phases are opposite each other.
Since the intensity of the wave in each space is relatively large whenever the wave forms
a belly, such a phase difference indicates that periodical intensity exchange
between the $q$- and $p$-component waves takes place
through their nonstatic evolutions.
This is the novel reciprocal optical phenomenon arisen on account of the
wave nonstaticity.
\\
\\
{\bf Keywords}: nonstatic light wave; wave function; measure of nonstaticity;
Gaussian state; quantum optics
\end{abstract}

\maketitle
\newpage

{\ \ \ } \\
{\bf 1. INTRODUCTION \vspace{0.2cm}}
\\
If the characteristics of a medium vary in time by external perturbations,
quantum light waves propagating through it may exhibit nonstatic
properties \cite{vat3,qpi,ivo,avdp,jap,bcn,1608,sab}.
Then, the shapes of the waves would be modified through the change of parameters.
The dissipation and amplification of the wave amplitudes are also classified as the
phenomena of wave nonstaticity.
The light waves in such cases are usually described by
time-dependent Hamiltonians, where their mathematical treatment is somewhat complicated.

Owing to the temporal and/or spatial variations of
electromagnetic parameters in media, nonstatic quantum waves
exhibit many novel physical properties that are absent in common light waves.
As a noticeable consequence of wave nonstaticity, highly rapid periodic
or arbitrary modulations of the wave phases and amplitudes
are possible under appropriate conditions \cite{1608,hd10,me,tms,wm2,wm3}.
Such ultrafast modulations at a weak
photon level can be applied to a time-resolved optical
heterodyne detection in nano quantum dots \cite{1608,hd10}.
On one hand, temporal modulation of a driving electromagnetic field
can be used in enhancing the entanglement between a microwave mode and a mechanical
resonator \cite{me,tms}.
Another main consequence achieved through wave nonstaticity is a frequency
shift \cite{ufr,sab} in subwavelength optics. Frequency shifts are potential tools for
producing millimeter-waves and terahertz signals
whose generations can hardly be realized by other technological means.

In a previous work \cite{nwh}, we showed a notable
feature in optics, which is that nonstatic waves can also appear even in a
static environment, i.e., without changes of the parameter values in media.
The properties of such nonstatic waves have been studied in a rigorous way
from the fundamental quantum-mechanical point of view in that work.
Through this, we
confirmed that the width of the waves varies periodically in time as a consequence of
their nonstaticity.
The related quantitative measure of nonstaticity, resulting from the modification of
the waveform, was defined.
Further, the mechanism of wave expansion and collapse was elucidated in Ref. \cite{gnb}.

However, the above mentioned research was in principle confined in $q$-quadrature space only.
In this work, we will investigate such nonstatic waves especially on their $p$-quadrature space
characteristics.
The behavior of nonstatic waves described in $p$-space will be compared  to that in $q$-space;
through this, we clarify how they are mutually connected,
as well as demonstrate the differences and similarities between them.
A rigorous analysis of the relation between the two conjugate space evolutions of
the nonstatic waves is necessary in understanding the peculiar wave behavior
caused by its nonstaticity as a whole.
Our analyses will be carried out separately for the Fock state waves and
the Gaussian ones.
\\
\\
{\bf 2. RESULTS AND DISCUSSION \vspace{0.2cm}} \\
{\bf 2.1. Nonstatic Waves in the Fock States \vspace{0.0cm}} \\
We investigate the properties of nonstatic waves in the Fock states in
a static environment through the $p$-space analysis in this section.
As a preliminary step
for understanding the theory of wave nonstaticity along this line,
the readers may need to know
its previous research consequence in $q$-quadrature space, reported in Ref. \cite{nwh}.
For the convenience of readers,
the research of Ref. \cite{nwh} including its outcome is briefly introduced in Appendix A.

Let us consider a medium where the electric permittivity $\epsilon$ and the magnetic
permeability $\mu$ do not vary in time, whereas the electric conductivity $\sigma$ is zero.
The vector potential relevant to the light-wave propagation
in that medium can be written as ${\bf A}({\bf r},t) = {\bf u}({\bf r})q(t)$,
where ${\bf r}$ is a position in three dimensions.
Whereas ${\bf u}({\bf r})$ follows position boundary conditions,
the time function $q(t)$ exhibits an oscillatory behavior.
In order to describe the light waves from quantum mechanical point of view,
we need to change $q(t)$ into an operator $\hat{q}$.
Then, the waves that propagate through the static medium is described by a
simple  Hamiltonian of the form
$\hat{H}= {\hat{p}^2}/{(2\epsilon)} + \epsilon \omega^2 \hat{q}^2 /2$,
where $\hat{p}$ is the conjugate variable of $\hat{q}$, which is defined as
$\hat{p}= -i\hbar \p/\p q$.
According to a previous report \cite{nwh}, quantum waves
that have nonstatic properties in $q$-space can appear even in this
static situation as mentioned in the introductory part.

The nonstatic waves propagating in the
static environment can be represented in terms of a time function of the form
\be
W(t) = W_{\rm R}(t) + i W_{\rm I}(t),  \label{1}
\ee
where $W_{\rm R}(t)$ and $W_{\rm I}(t)$ are its real and imaginary parts, respectively
(i.e., both $W_{\rm R}(t)$ and $W_{\rm I}(t)$ are real).
The formulae of the real and imaginary parts are given by \cite{nwh}
\be
W_{\rm R}(t) = \f{\epsilon \omega}{\hbar f(t)},~~~~~~~~~~~~~~~~~~  \label{} \\
W_{\rm I}(t) = -\f{\epsilon \dot{f}(t)}{2\hbar f(t)},  \label{3}
\ee
where
\ba
& &f(t) = A \sin^2 \tilde{\varphi}(t)+ B \cos^2 \tilde{\varphi}(t) +
C \sin [2\tilde{\varphi}(t)], \label{4} \\
& &\tilde{\varphi}(t)=\omega (t-t_0) +\varphi, \label{5}
\ea
with a constant phase $\varphi$ and a constant time $t_0$.
In Eq. (\ref{4}), $A$, $B$, and $C$ are real values that obey
the conditions $AB-C^2 = 1$ and $AB \geq 1$.
We note that Eq. (\ref{4}) is a general solution of the nonlinear equation
\cite{nwh}
\be
\ddot{f} - {(\dot{f})^2}/({2f}) + 2\omega^2 \left(f- {1}/{f}\right) =0. \label{}
\ee

The wave functions in $p$-space, which exhibit nonstatic properties, may also be represented
in terms of the time function given in Eq. (\ref{1}).
An exact evaluation of the wave functions for such waves in the Fock state results in
(see Appendix B)
\be
\langle p |{\psi}_n(t) \rangle = \langle p |\phi_n(t) \rangle
 \exp [i\gamma_n(t) ], \label{6}
\ee
where
\ba
\langle p |\phi_n \rangle &=& (-i)^n
\left({\f{W_{\rm R}(t)}{\pi\hbar^2}}\right)^{1/4} \sqrt{\f{[W^*(t)]^n}{[W(t)]^{n+1}}} \f{1}{\sqrt{2^n
n!}} \nonumber \\
& &\times H_n \left( \sqrt{\f{W_{\rm R}(t)}{W_{\rm R}^2(t) +W_{\rm I}^2(t)}} \f{p}{\hbar} \right) \exp \left[
- \f{W_p(t)}{2} p^2 \right], \label{7} \\
\gamma_n(t) &=& {-\omega (n+1/2) \int_{t_0}^t f^{-1} (t') dt'} + \gamma_n (t_0),
 \label{8}
\ea
with
\be
W_p(t) = \f{W_{\rm R}(t)-i W_{\rm I}(t)}{\hbar^2 [W_{\rm R}^2(t)+W_{\rm I}^2(t)]}
\equiv W_{p,{\rm R}}(t)+i W_{p,{\rm I}}(t). \label{9}
\ee
Here, $H_{n}$ are $n$th order Hermite polynomials, whereas
$W_{p,{\rm R}}(t)$ and $W_{p,{\rm I}}(t)$ are real and imaginary parts of $W_p(t)$, respectively.
We can also represent Eq. (\ref{9}) simply as $W_p(t) = {1}/{[\hbar^2 W(t)]}$.

\begin{figure}
\centering
\includegraphics[keepaspectratio=true]{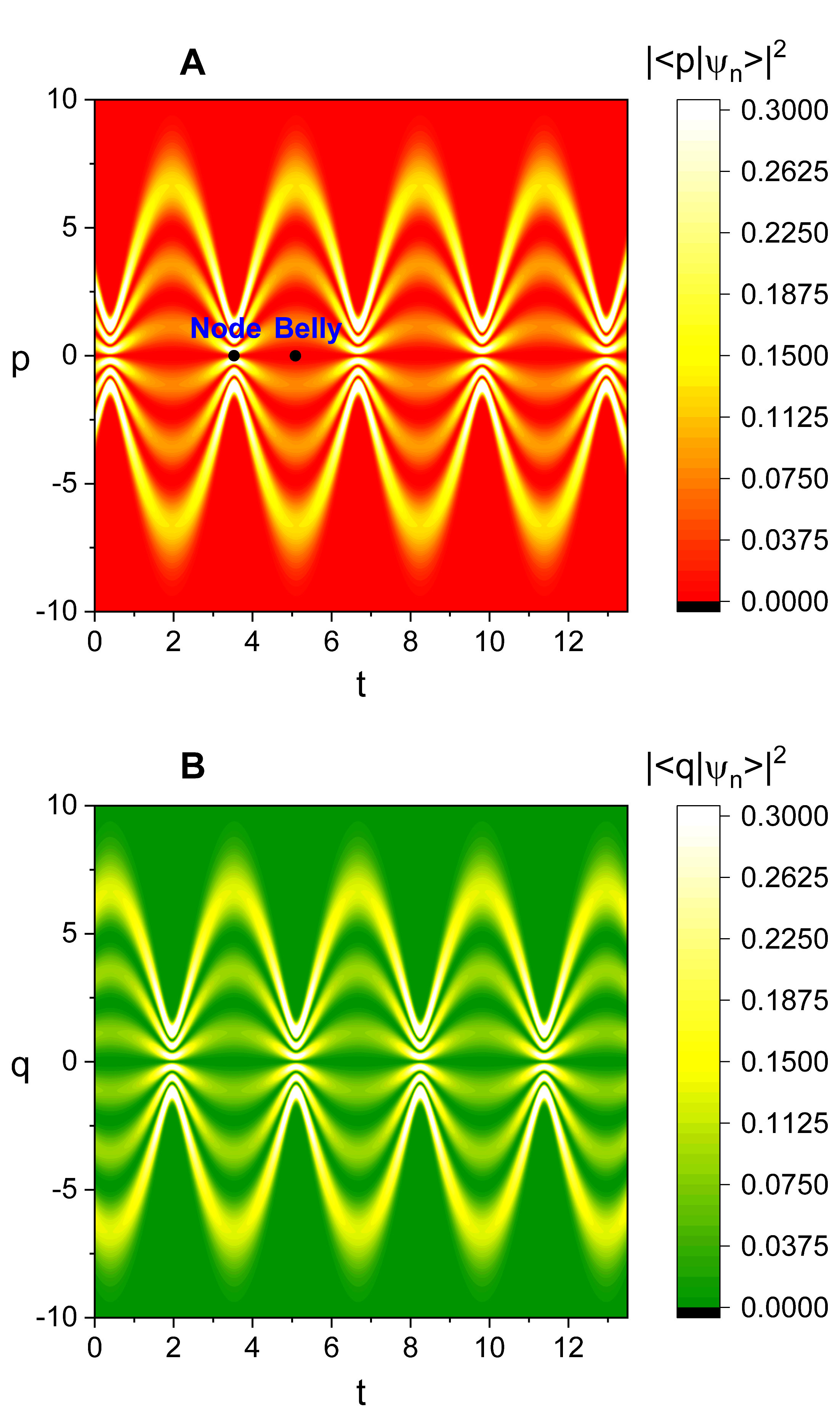}
\caption{\label{Fig1} {\textsf A} is the density plot of the probability density
in $p$-space, $|\langle p |{\psi}_n(t) \rangle|^2$, as a function of $p$ and $t$.
For comparison purposes, we have also represented the probability density
in $q$-space, $|\langle q |{\psi}_n(t) \rangle|^2$, in {\textsf B} as a function of $q$ and $t$.
The values that we used are $A=1$, $B =5$, $C =2$, $\omega = 1$, $n=5$, $\hbar = 1$,
$\epsilon=1$, $t_0 = 0$, and $\varphi = 0$. All values are taken to be dimensionless
for convenience; this convention will also be used in all subsequent figures.
}
\end{figure}

Now it is possible to analyze quantum optical phenomena associated with the nonstatic waves
on the basis of the wave functions given in Eq. (\ref{6}),
Figure 1 shows the temporal evolution of the probability density which is
defined as $|\langle p |{\psi}_n(t) \rangle|^2$.
We see from Fig. 1({\textsf A}) that there are bellies and nodes in the $p$-space wave
evolution like the $q$-space evolution represented in Fig. 1({\textsf B}).
While the period of such evolutions is $\pi/\omega$ for both
$|\langle p |\phi_n \rangle|^2$ and $|\langle q |\phi_n \rangle|^2$,
the corresponding wave phases associated with wave nonstaticity
are different from each other.
We confirm, from the comparison of Fig. 1({\textsf A}) and Fig. 1({\textsf B}),
that such a phase difference between the two probability densities is $\pi$, i.e.,
the phases are opposite each other.
Notice that the phase of nonstatic evolution that we use here is the one
that emerges due to nonstatic characteristics of the wave: this concept of the phase
is essentially different
from the generally used quantum phase which is composed of the dynamical
and geometric phases as shown, for example, in Ref. \cite{igp}.
Since the wave intensity is strong at the belly and weak at the node,
the $\pi$ difference between the $q$- and $p$-space wave phases
implies the exchange of the wave intensity between the two conjugate components of the wave. 

As stated in Appendix A, it was shown in the previous work \cite{nwh} that the non-zero
value of $W_{\rm I}(t)$ is responsible for the appearance of the nonstatic
properties of the waves in $q$-space.
In addition, the quantitative measure of nonstaticity in $q$-space was defined
in the same reference as the root-mean-square (RMS) value of $W_{\rm I}(t)/W_{\rm R}(t)$.

\begin{figure}
\centering
\includegraphics[keepaspectratio=true]{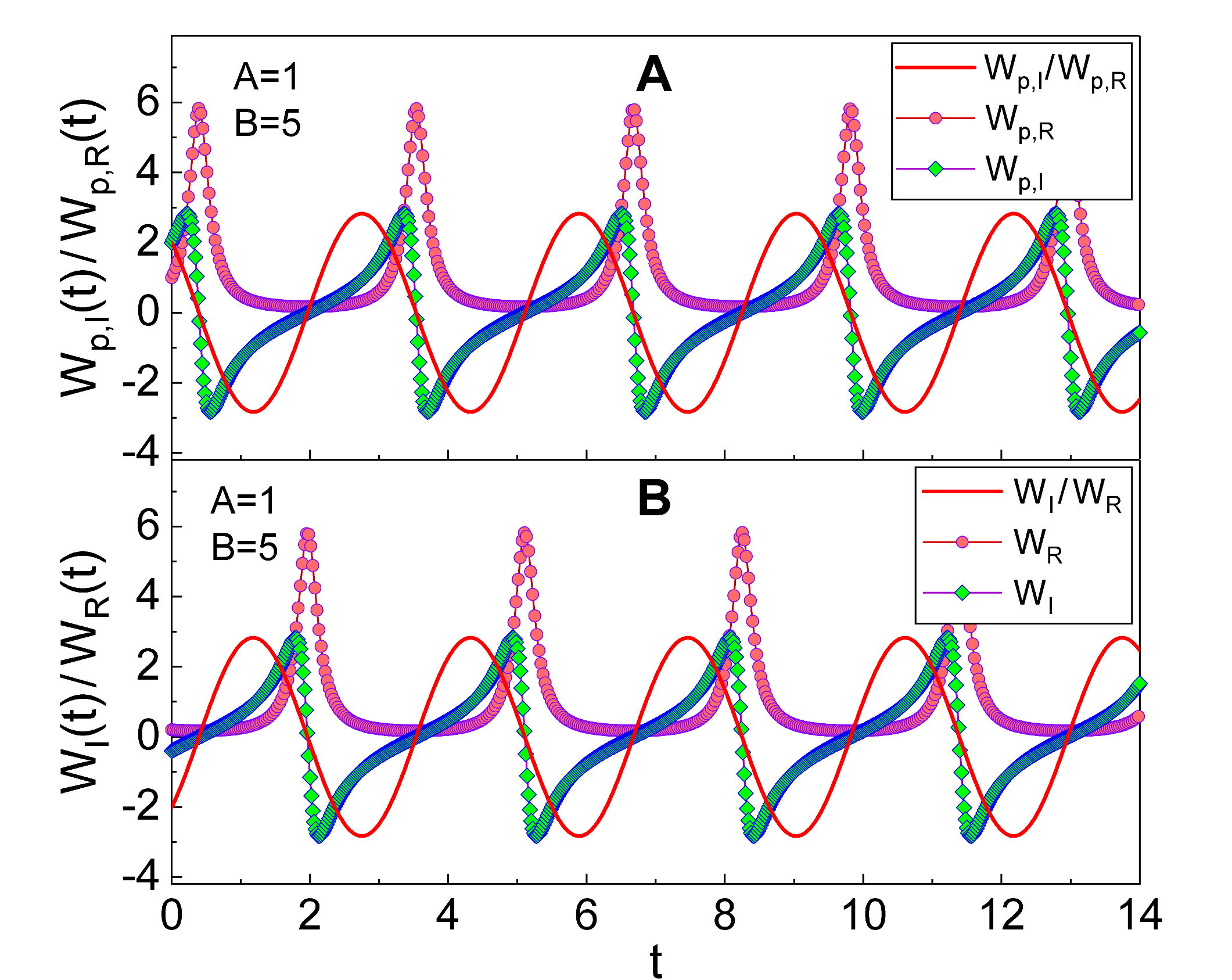}
\caption{\label{Fig2} {\textsf A} is the evolution of $W_{p,{\rm I}}/W_{p,{\rm R}}$
and its components $W_{p,{\rm R}}$ and $W_{p,{\rm I}}$,
relevant to $p$-space, with the choice of $A=1$, $B =5$, and $C =2$.
{\textsf B} is the evolution of $W_{\rm I}/W_{\rm R}$ and its components
relevant to $q$-space (comparison purpose).
The values that we have taken are $\omega = 1$, $\hbar = 1$,
$\epsilon=1$, $t_0 = 0$, and $\varphi = 0$.
}
\end{figure}

For the case of the wave description in $p$-space, the associated wave nonstaticity originates
from the non-zero value of $W_{p,{\rm I}}$.
Hence, the nonstaticity measure in $p$-space can also be defined
as the RMS value of $W_{p,{\rm I}}/W_{p,{\rm R}}$.
We see from Fig. 2 that the evolution of $W_{p,{\rm I}}/W_{p,{\rm R}}$ exhibits sinusoidal behavior
as that of $W_{\rm I}/W_{\rm R}$. However, there is a phase difference $\pi$ between them.
This difference is responsible for the phase difference between the evolutions of
$|\langle p |\psi_n \rangle|^2$ and $|\langle q |\psi_n \rangle|^2$.
From a minor evaluation, we can easily confirm that
\be
\f{W_{p,{\rm I}}}{W_{p,{\rm R}}} = \f{1}{2} \sqrt{(A+B)^2-4}\cos [2\tilde{\varphi}(t) +\delta],
 \label{9-1}
\ee
where $\delta = {\rm atan}(2C,B-A)$.
Here, $\theta \equiv {\rm atan}(x,y)$ is the two-arguments inverse function of $\tan\theta = y/x$,
which is defined in the range $0 \leq \theta < 2\pi$.
We have depicted the temporal evolution of $W_{p,{\rm I}}/W_{p,{\rm R}}$ and its components
$W_{p,{\rm R}}$ and $W_{p,{\rm I}}$
in Fig. 2. While $W_{p,{\rm I}}/W_{p,{\rm R}}$ varies sinusoidally,
$W_{p,{\rm R}}$ and $W_{p,{\rm I}}$ vary abruptly at certain instants of time
where the $p$-space probability density constitutes nodes.
By taking the RMS value of Eq. (\ref{9-1}) for a cycle,
we have the measure of nonstaticity in $p$-space as
\be
D_{p,{\rm F}} = \f{\sqrt{(A+B)^2-4}}{2\sqrt{2}}. \label{n9-1}
\ee
This is exactly the same as the measure of nonstaticity in $q$-space, which was
previously evaluated in Ref. \cite{nwh}.
Thus, the definition of the measure of nonstaticity shown above can be generally used
irrespective of the given space.
For instance, we confirm that the measure of nonstaticity for the wave
given in Fig. 2 is $2.0$ from a simple calculation using $A=1$ and $B=5$.
\\
\\
{\bf 2.2. Nonstatic Waves in the Gaussian States \vspace{0.0cm}} \\
Our theory for the evolution of nonstatic waves can be extended to a more general case
which is the Gaussian wave.
To see the nonstatic properties of a Gaussian wave that evolves in
a static environment, we take an initial waveform as
\be
\langle q |\psi(0) \rangle =  \sqrt[4]{\f{K_{\rm R}}{\pi}} e^{-\f{1}{2}K(q-\xi)^2}, \label{10}
\ee
where $\xi$ is a displacement and $K = K_{\rm R}+i K_{\rm I}$. Here, $K_{\rm R}$
and $K_{\rm I}$ mean real and imaginary parts, respectively.
The existence of $K_{\rm I}$ is responsible for the nonstatic evolution
of the wave in this case \cite{nwh}.
To see the evolution of this Gaussian quantum wave in $p$-space, it is necessary to evaluate
the wave function $\langle p |\psi(t) \rangle$ at an arbitrary time $t$ from Eq. (\ref{10}).
\begin{figure}
\centering
\includegraphics[keepaspectratio=true]{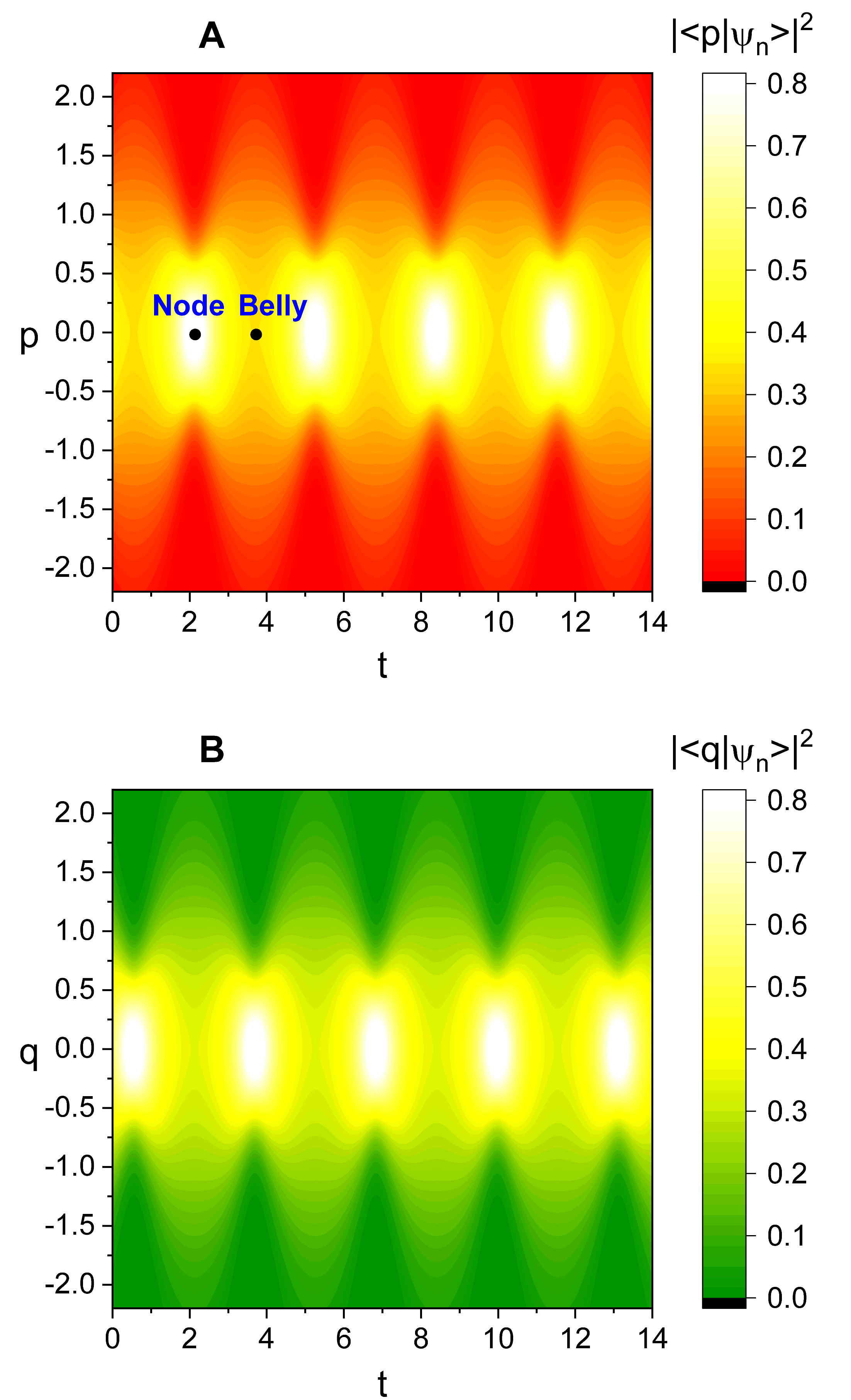}
\caption{\label{Fig3} {\textsf A} is the density plot of the probability density
$|\langle p |\phi_n \rangle|^2$ given as a function of $p$ and $t$.
{\textsf B} is the density plot of $|\langle q |\phi_n \rangle|^2$ given as a
function of $q$ and $t$ (comparison purpose).
The values that we have taken are $K_{\rm R} =1$, $K_{\rm I} =1$, $A=1$,
$B =5$, $C =2$, $\xi=0$, $\omega = 1$, $\hbar = 1$, $\epsilon=1$, $t_0 = 0$, and $\varphi = 0$.
}
\end{figure}
We have provided the method of evaluating the analytical formula
for such a wave function in Appendix C and the result is given by
\be
\langle p |\psi(t) \rangle = \f{N(t)}{\sqrt{\hbar {\mathcal W}(t)}}
\exp \bigg( -\f{{\mathcal W}_p(t)}{2}[p+ i\hbar R(t)]^2 \bigg),
 \label{11}
\ee
where
\ba
& &N(t) = \bigg( \f{W_{\rm R}(0)W_{\rm R}(t)}{\pi} \bigg)^{1/4}
\bigg( \f{2K_{\rm R}^{1/2}}{ g(t)\exp[-2i\Theta(t) ]} \bigg)^{1/2}
 \exp\bigg[-\f{1}{2}\Big(K \xi^2 +i\Theta(t) \Big) \bigg]
\nonumber \\
& &~~~~~~~~~~~\times \exp \bigg[ \f{K^2 \xi^2}{K+W^*(0)} \bigg( \f{1}{2}
- \f{W_{\rm R}(0)}{g(t)} \bigg) \bigg], \label{12} \\
& &{\mathcal W}(t) = W(t)+ \f{2W_{\rm R}(t)[K-W(0)]}{g(t)}
\equiv {\mathcal W}_{\rm R}(t)+i {\mathcal W}_{\rm I}(t), \label{13} \\
& &{\mathcal W}_p(t) 
= \f{{\mathcal W}_{\rm R}(t)-i {\mathcal W}_{\rm I}(t)}{\hbar^2
[{\mathcal W}_{\rm R}^2(t)+{\mathcal W}_{\rm I}^2(t)]}
\equiv {\mathcal W}_{p,{\rm R}}(t)+i {\mathcal W}_{p,{\rm I}}(t),
 \label{14} \\
& &R(t) = \f{2K \xi \sqrt{W_{\rm R}(0)W_{\rm R}(t)} }{g(t)  \exp [-i\Theta(t)]}, \label{15}
\ea
with
\ba
\Theta(t) &=& \omega \int_0^t f^{-1}(t')dt' , \label{16} \\
g(t) &=& [K+W^*(0)] \exp [2i\Theta(t) ]-[K-W(0)]. \label{17}
\ea

\begin{figure}
\centering
\includegraphics[keepaspectratio=true]{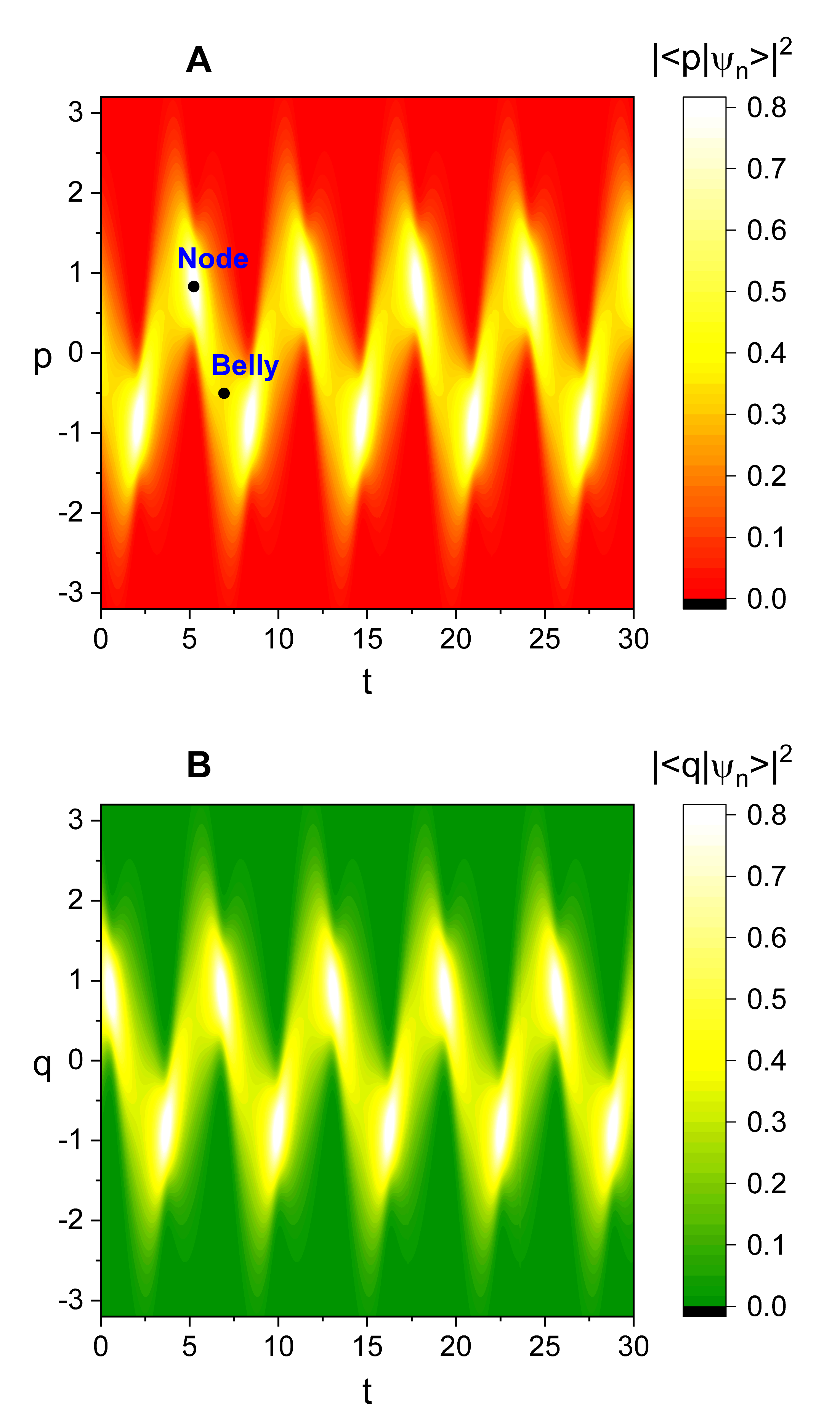}
\caption{\label{Fig4} This is the same as Fig. 3, but with the choice of
$\xi = 1$.
}
\end{figure}

\begin{figure}
\centering
\includegraphics[keepaspectratio=true]{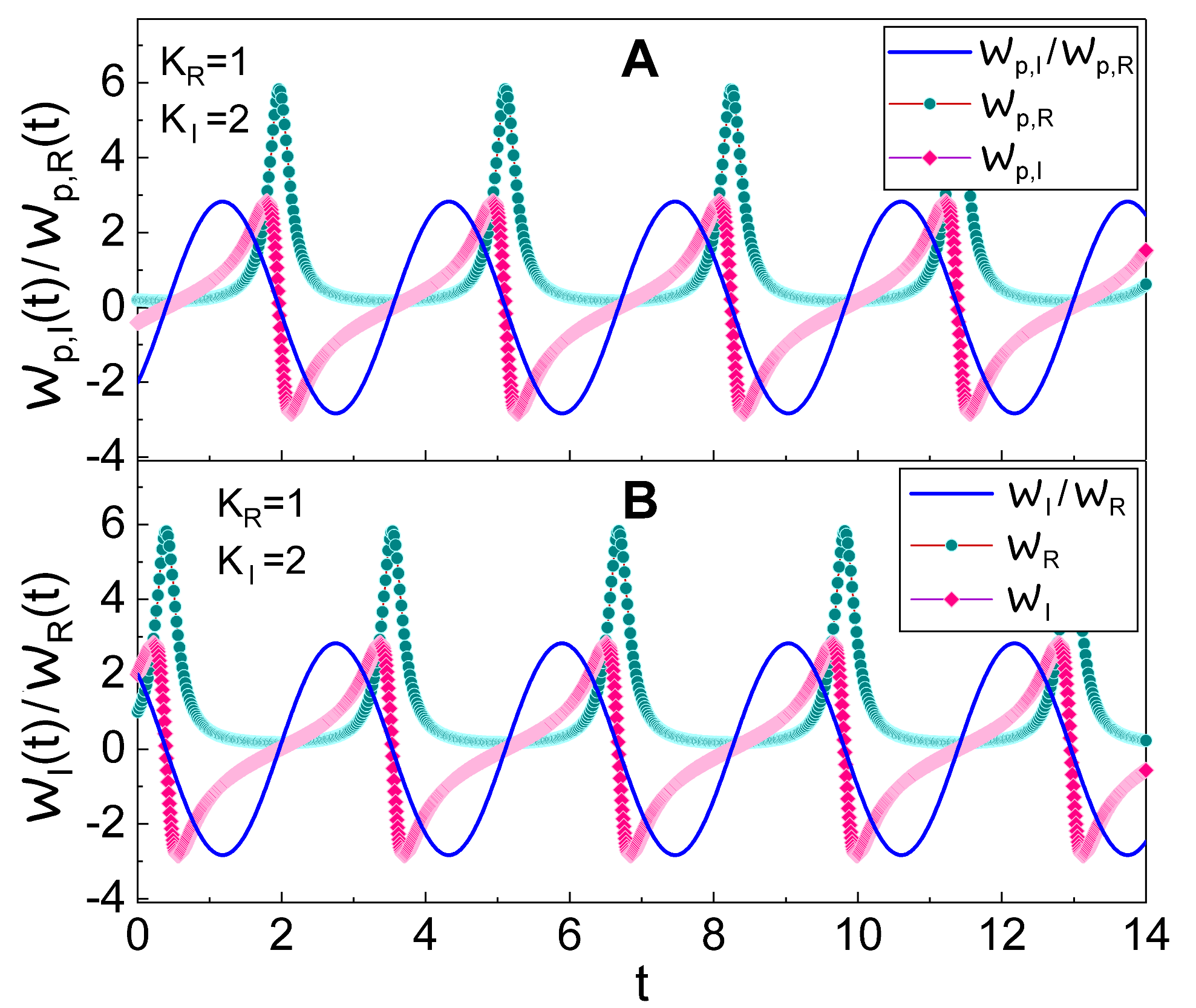}
\caption{\label{Fig5} {\textsf A} is the evolution of ${\mathcal W}_{p,{\rm I}}/{\mathcal W}_{p,{\rm R}}$
with the components ${\mathcal W}_{p,{\rm R}}$ and ${\mathcal W}_{p,{\rm I}}$,
where $K_{\rm R} =1$ and $K_{\rm I} =2$.
{\textsf B} is the evolution of ${\mathcal W}_{\rm I}/{\mathcal W}_{\rm R}$ and
its relevant components (comparison purpose).
The values that we have used are  $A=1$,
$B =5$, $C =2$, $\omega = 1$, $\hbar = 1$, $\epsilon=1$, $t_0 = 0$, and $\varphi = 0$.
}
\end{figure}

We have illustrated the probability density that corresponds to Eq. (\ref{11}) in Figs. 3 and 4.
Figure 3 is the case where the displacement $\xi$ is zero, whereas
Fig. 4 corresponds to the case of the displaced Gaussian wave.
We see from Fig. 3 that the width of the probability density varies periodically over time with the
period 3.14 ($=\pi/\omega$), which is the same period
as that of the Fock-state wave functions that we have already seen.
The comparison of Figs. 3({\textsf A}) and 3({\textsf B}) shows that the phase difference between
$|\langle p |\psi(t) \rangle|^2$ and $|\langle q |\psi(t) \rangle|^2$ is $\pi$,
which is also the same as that between the Fock-state probability densities
$|\langle p |\psi_n(t) \rangle|^2$ and $|\langle q |\psi_n(t) \rangle|^2$.

On the other hand, the period of wave evolution (oscillation) in the case
of Fig. 4 is not 3.14 
but 6.28 ($=2\pi/\omega$).
This means that the wave evolves a one-cycle ($2\pi$ rad) during $T=6.28$.
Based on this, we can conclude that the phase difference between
Figs. 4({\textsf A}) and 4({\textsf B}) is $\pi/2$ which is different from the previous cases.
The difference of the period in this case from that in the previous cases is
responsible for such an inconsistency
in the phase differences ($p$-space wave phase minus $q$-space wave phase) between them.
However, if we neglect the oscillation of the Gaussian wave in Fig. 4, the period of its evolution
reduces to $\pi$ and  as a consequence, the phase difference between
$|\langle p |\psi(t) \rangle|^2$ and $|\langle q |\psi(t) \rangle|^2$ becomes $\pi$ which is
identical to the previous two cases (Figs. 1 and 3).

The displaced Gaussian wave that can be seen
from Fig. 4 oscillates back and forth like a classical state.
However, the shape of the wave varies in an abnormal manner in time
due to its nonstatic properties.
The value of the probability density is highly peaked
whenever the value of $p$ becomes one of certain two values (one is plus and the other is minus)
just after the turning points of the oscillation in $p$-space.
Such peaks are in fact nodes as designated in the figure.
Because the Gaussian nonstatic wave also exhibits node and belly in turn regularly,
the intensity exchange between the $q$- and $p$-space waves occurs.

We can define the measure of nonstaticity for the Gaussian wave in a similar manner as that of the
previous section. It is the RMS value of
${\mathcal W}_{p,{\rm I}}(t)/{\mathcal W}_{p,{\rm R}}(t)$.
The temporal evolution of ${\mathcal W}_{p,{\rm I}}(t)/{\mathcal W}_{p,{\rm R}}(t)$
has been represented in Fig. 5 with its comparison to the evolution of the counterpart $q$-space
value ${\mathcal W}_{\rm I}(t)/{\mathcal W}_{\rm R}(t)$.
The amplitude of both ${\mathcal W}_{p,{\rm I}}(t)/{\mathcal W}_{p,{\rm R}}(t)$
and ${\mathcal W}_{\rm I}(t)/{\mathcal W}_{\rm R}(t)$ in Fig. 5 is 2.83.
From this, the corresponding measure of nonstaticity is 2.0 and this value is the same as that
of the wave shown in Fig. 2.
We can confirm that the patterns of the evolutions of
${\mathcal W}_{p,{\rm R}}(t)$ and ${\mathcal W}_{p,{\rm I}}(t)$ given in Fig. 5
are the same as those of ${W}_{p,{\rm R}}(t)$ and ${W}_{p,{\rm I}}(t)$
given in Fig. 2, respectively, except for the phases in their evolutions.
From this, we can conclude that if the measure of nonstaticity
is the same as each other, the patterns of their temporal evolutions are also the same.
On the other hand, we can confirm by comparing panels of Fig. 3 and Fig. 6
in Ref. \cite{nwh} one another
that, if the measure of nonstaticity is different, the patterns of their evolutions are no longer
the same.
However, ${\mathcal W}_{p,{\rm I}}(t)/{\mathcal W}_{p,{\rm R}}(t)$
and ${W}_{p,{\rm I}}(t)/{W}_{p,{\rm R}}(t)$ always undergo sinusoidal evolution
in any case in the static environment.
\\
\\
{\bf 3. CONCLUSION \vspace{0.2cm}} \\
Through the extension of the research in $q$-space nonstatic-wave phenomena to its conjugate
$p$-space ones, we pursued better understanding of wave nonstaticity
including the interaction of the two component waves.
Our analysis was carried out purely on the basis of analytical methods,
where we did not use any approximation.

We have shown that bellies and nodes appear
in the $p$-space evolution of the Fock and Gaussian state quantum-waves
as a manifestation of their nonstaticity, like in the case of $q$-space evolution.
The evolving pattern of the $p$-space wave caused by its nonstaticity is
in general out of phase by $\pi$ with the $q$-space evolution of the wave.
This implies that the two wave components are reciprocally linked.
Because the wave intensity is large when it constitutes a belly, the wave in each space
gives and receives
intensity from the conjugate wave-component periodically.
If there is an initial displacement in the Gaussian wave, the wave in $p$-space
oscillates back and forth like the $q$-space wave.
This behavior very much resembles
classical waves. However, the waveform in such an oscillation
was altered due to the appearance of bellies and nodes.

Whenever the wave in $q$-space ($p$-space) poses a node, the uncertainty of $q$ ($p$) reduces
below its standard quantum level.
From this, we can conclude that the nonstatic wave treated here is a kind of squeezed state.
Several methods of generating squeezed states are known until now \cite{ssg1,ssg2,ssg3,ssg4,ssg5}.
Likewise, the nonstatic wave may also be generated by using the technique of
squeezed-state generation.
The nonstatic wave can be used in physical disciplines where the squeezed state
plays a major role, such as interferometers of gravitational-wave detection \cite{gwd1,gwd2,gwd3},
quantum information processing \cite{qip1,qip2,qip3}, and high-precision measurements \cite{hpm1,hpm2}.

It may be noticeable that nonstatic waves can arise
without temporal changes of the electromagnetic parameters in media.
However, we can never say that we know wave nonstaticity well if our related knowledge is limited
to only $q$--space behavior of the light waves.
The outcome of this research complements previous $q$-quadrature
analyses in this context \cite{nwh}.
Based on this research, we can outline the entire aspect of wave nonstaticity
including the integral connection between the $q$- and $p$-space wave behaviors.
\appendix
\section{\bf Previous Research Consequence (Ref. \cite{nwh})}
Here we summarize the research of Ref. \cite{nwh}, which belongs to the $q$-space wave-nonstaticity
in a static environment. General types of wave functions for Fock-state waves and a Gaussian wave
were established on the basis of the Schr\"{o}dinger equation. Such wave functions were represented
in terms of the time function $f(t)$ given in Eq. (\ref{4}) in the present work and showed nonstaticity
(hereafter, all referred equations in this appendix belong to the present work). According to the nonstaticity
of waves, periodical collapse and expansion of the $q$-space waves appeared.
It was shown, in case of the Fock states, that the emergence of the imaginary part $W_{\rm I}(t)$ in $W(t)$ that
appears in the exponential factor of wave functions (see Eq. (\ref{19}) in Appendix B
with Eq. (\ref{1})) is responsible for the nonstaticity in the wave.
Taking notice of this consequence, the measure of nonstaticity was defined as
the RMS value of the ratio $W_{\rm I}(t)/W_{\rm R}(t)$ where $W_{\rm R}(t)$ is the
real part of $W(t)$.
The same methodology but with different time functions was applied to the Gaussian state
in quantifying its nonstaticity.
It was shown that the effect of nonstaticity becomes significant as the nonstaticity measure grows.
\section{\bf Evaluation of the Fock-State Wave Functions in $p$-space}
The wave functions which exhibit nonstatic properties in $p$-space can be obtained from
the Fourier transformation of the wave functions in $q$-space.
The wave functions for the nonstatic waves in $q$-space, which propagate in a
static environment, are given by \cite{nwh}
\be
\langle q |\psi_n(t) \rangle = \langle q |\phi_n(t) \rangle
 \exp [i\gamma_n(t) ], \label{18}
\ee
where $\langle q |\phi_n \rangle$ are eigenstates of the form
\be
\langle q |\phi_n \rangle =
\left({\f{W_{\rm R}(t)}{\pi}}\right)^{1/4} \f{1}{\sqrt{2^n
n!}} H_n \left( \sqrt{W_{\rm R}(t)} q \right) \exp \left[
- \f{W(t)}{2} q^2 \right]. \label{19}
\ee
Let us carry out the Fourier transformation of these waves, such that
\be
\langle {p}|\psi_n (t)\rangle  = \frac{1}{\sqrt{2\pi \hbar}}
\int_{-\infty}^{\infty} \langle q|\psi_n (t)\rangle
e^{-i {p}{q}/\hbar} d q.  \label{20}
\ee
By evaluating the above equation using Eqs. (\ref{18}) and (\ref{19}) straightforwardly,
we easily have the $p$-space wave functions, which are given in
Eq. (\ref{6}) with Eqs. (\ref{7})-(\ref{9}) in the text.
\section{The Gaussian Wave in $p$-space}
The Gaussian wave in $q$-space, that exhibits nonstatic properties, was suggested in Ref. \cite{nwh}.
From that reference, the corresponding wave function is given by
\ba
& &\langle q |\psi(t) \rangle = N(t) \exp \Bigg[ - \f{\mathcal{W}(t)}{2}q^2
+ R(t) q \Bigg], \label{21}
\ea
where $ N(t)$, $\mathcal{W}(t)$, and $R(t)$ are defined in the text.
From the Fourier transformation of this,
we have
the exact wave function for the Gaussian wave in $p$-space, which is given in Eq. (\ref{11}) with
Eqs. (\ref{12})-(\ref{17}).
\\
\\
{\bf Acknowledgements \vspace{0.0cm}} \\
This work is focused on the behavior of $p$-quadrature wave functions,
but we also provide the graphics of the physical quantities associated with
$q$-quadrature for comparison purposes.
The graphics that belong to $q$-quadrature
were depicted using the analytical evaluations given in Ref. \cite{nwh}. \\



\begin{references}

\bibitem{qpi} Dodonov, V.V., Klimov, A.B., Nikonov, D.E.:
Quantum phenomena in nonstationary media.
{Phys. Rev. A} {\bf 47}(5), 4422-4429 (1993).
https://doi.org/10.1103/physreva.47.4422

\bibitem{vat3} Akhmanov, S.A., Sukhorukov, A.P., Chirkin, A.S.:
Nonstationary phenomena and space-time analogy in nonlinear optics.
{Soviet Phys. JETP} {\bf 28}(4), 748-757 (1969).

\bibitem{ivo} Vorgul, I.:
On Maxwell's equations in non-stationary media.
{Phil. Trans. R. Soc. A} {\bf 366}(1871), 1781-1788 (2008).
https://doi.org/10.1098/rsta.2007.2186

\bibitem{avdp} Dodonov, A.V.:
Photon creation from vacuum and interactions engineering in nonstationary circuit QED.
{J. Phys.: Conf. Ser.} {\bf 161}(1), 012029 (2009).
https://doi.org/10.1088/1742-6596/161/1/012029

\bibitem{jap} Porti, J.A., Salinas, A., Morente, J.A., Rodriguez-Sola, M., Nerukh, A.G.:
Time-varying electromagnetic-media modelling with TLM method.
{Electron. Lett.} {\bf 39}(6), 505-507 (2003).
https://doi.org/10.1049/el:20030390

\bibitem{bcn} Bakunov, M.I., Grachev, I.S.:
Energetics of electromagnetic wave transformation in a time-varying magnetoplasma medium.
{Phys. Rev. E} {\bf 65}(3), 036405 (2002).
https://doi.org/10.1103/PhysRevE.65.036405

\bibitem{sab} Shvartsburg, A., Petite, G.:
Instantaneous optics of ultrashort broadband pulses and rapidly varying media.
{Prog. Opt.} {\bf 44}, 143-214 (2002).
https://doi.org/10.1016/S0079-6638(02)80016-6

\bibitem{1608} Moody, G., McDonald, C., Feldman, A., Harvey, T., Mirin, R.P., Silverman, K.L.:
Quadrature demodulation of a quantum dot optical response to faint light fields.
{Optica} {\bf 3}(12), 1397-1403 (2016).
https://doi.org/10.1364/OPTICA.3.001397

\bibitem{wm2} Bogatov, A.P., D'yachkov, N.V., Drakin, A.E., Gushchik, T.I.:
Amplitude and phase modulation of radiation in a travelling-wave amplifier based on a laser diode.
{Quantum Electron.} {\bf 43}(8) 699-705 (2013).
https://doi.org/10.1070/QE2013v043n08ABEH015166

\bibitem{wm3} Waarts, R.G., Friesem, A.A., Hefetz, Y.:
Frequency-modulated to amplitude-modulated signal conversion by a Brillouin-induced phase change
in single-mode fibers.
{Opt. Lett.} {\bf 13}(2), 152-154 (1988).
https://doi.org/10.1364/OL.13.000152

\bibitem{hd10} Liu, B., Li, X., Zhang, Y., Xin, X., Yu, J.:
Probabilistic shaping for ROF system with heterodyne coherent detection.
APL Photonics {\bf 2}(5), 056104 (2017).
https://doi.org/10.1063/1.4981393

\bibitem{me} Mari, A., Eisert, J.:
Opto- and electro-mechanical entanglement improved by modulation. New J. Phys. {\bf 14}, 075014 (2012).
https://doi.org/10.1088/1367-2630/14/7/075014

\bibitem{tms} Averchenko, V., Sych, D., Schunk, G., Vogl, U., Marquardt, C., Leuchs, G.:
Temporal shaping of single photons enabled by entanglement.
Phys. Rev. A {\bf 96}(4), 043822 (2017).
https://doi.org/10.1103/PhysRevA.96.043822

\bibitem{ufr} Savage Jr., R.L., Joshi, C., Mori, W.B.:
Frequency upconversion of electromagnetic radiation upon transmission into an ionization front.
{Phys. Rev. Lett.} {\bf 68}(7), 946-949 (1992).
https://doi.org/10.1103/PhysRevLett.68.946

\bibitem{nwh} Choi, J.R.:
On the possible emergence of nonstatic quantum waves in a static environment.
Nonlinear Dyn. {\bf 103}(3), 2783-2792 (2021).
https://doi.org/10.1007/s11071-021-06222-8

\bibitem{gnb} Choi, J.R.:
Characteristics of nonstatic quantum light waves: the principle for wave expansion and collapse.
{Photonics} {\bf 8}(5), 158 (2021).
https://doi.org/10.3390/photonics8050158

\bibitem{igp} Choi, J.R.:
Quadrature squeezing and geometric-phase oscillations in nano-optics.
Nanomaterials {\bf 10}(7), 1391 (2020). https://doi.org/10.3390/nano10071391

\bibitem{ssg1} Han, Y., Wen, X., Liu, J., He, J., Wang, J.:
Generation of polarization squeezed light with an optical parametric amplifier at 795 nm.
Opt. Commun. {\bf 416}, 1-4 (2018). https://doi.org/10.1016/j.optcom.2018.01.038

\bibitem{ssg2} Ma, L., Guo, H., Sun, H., Liu, K., Su, B., Gao, J.:
Generation of squeezed states of light in arbitrary complex amplitude transverse distribution.
Photonics Res. {\bf 8}(9), 1422-1427 (2020). https://doi.org/10.1364/PRJ.388956

\bibitem{ssg3} Li, Y., Xiao, M.:
Generation and applications of amplitude-squeezed states of light from semiconductor diode lasers.
Opt. Express {\bf 2}(3), 110-117 (1998). https://doi.org/10.1364/OE.2.000110

\bibitem{ssg4} Raizen, M.G., Orozco, L.A., Xiao, M., Boyd, T.L., Kimble, H.J.:
Squeezed-state generation by the normal modes of a coupled system.
Phys. Rev. Lett. {\bf 59}(2), 198-201 (1987). https://doi.org/10.1103/PhysRevLett.59.198

\bibitem{ssg5} Yurke, B.:
Use of cavities in squeezed-state generation.
Phys. Rev. A {\bf 29}(1), 408-410 (1984). https://doi.org/10.1103/PhysRevA.29.408

\bibitem{gwd1} Grote, H., Danzmann, K., Dooley, K.L., Schnabel, R., Slutsky, J., Vahlbruch, H.:
First long-term application of squeezed states of light in a gravitational-wave observatory.
Phys. Rev. Lett. {\bf 110}(18), 181101 (2013). https://doi.org/10.1103/PhysRevLett.110.181101

\bibitem{gwd2} Aasi, J. et al. (The LIGO Scientific Collaboration):
Enhanced sensitivity of the LIGO gravitational wave detector by using squeezed states of light.
Nat. Photonics {\bf 7}(8), 613-619 (2013). https://doi.org/10.1038/nphoton.2013.177

\bibitem{gwd3} Abadie, J. et al. (The LIGO Scientific Collaboration):
A gravitational wave observatory operating beyond the quantum shot-noise limit.
Nat. Phys. {\bf 7}, 962-965 (2011). https://doi.org/10.1038/nphys2083

\bibitem{qip1} Appel, J., Figueroa, E., Korystov, D., Lobino, M., Lvovsky, A.I.:
Quantum memory for squeezed light. Phys. Rev. Lett. {\bf 100}(9), 093602 (2008).
https://doi.org/10.1103/PhysRevLett.100.093602

\bibitem{qip2} Honda, K., Akamatsu, D., Arikawa, M., Yokoi, Y., Akiba, K.,
Nagatsuka, S., Tanimura, T., Furusawa, A., Kozuma, M.:
Storage and retrieval of a squeezed vacuum.
Phys. Rev. Lett. {\bf 100}(9), 093601 (2008). https://doi.org/10.1103/PhysRevLett.100.093601

\bibitem{qip3} H\'{e}tet, G., Buchler, B.C., Gl\"{o}ckl, O., Hsu, M.T.L.,
 Akulshin, A.M., Bachor, H.?A., Lam, P.K.:
Delay of squeezing and entanglement using electromagnetically induced transparency in
a vapour cell. Opt. Express {\bf 16}(10), 7369-7381 (2008).
https://doi.org/10.1364/OE.16.007369

\bibitem{hpm1} Liu, K., Cai, C., Li, J., Ma, L., Sun, H., Gao, J.:
Squeezing-enhanced rotating-angle measurement beyond the quantum limit.
Appl. Phys. Lett. {\bf 113}(26), 261103 (2018). https://doi.org/10.1063/1.5066028

\bibitem{hpm2} Smithey, D.T., Beck, M., Raymer, M.G., Faridani, A.:
Measurement of the Wigner distribution and the density matrix
of a light mode using optical homodyne tomography: application to squeezed
states and the vacuum.
Phys. Rev. Lett. {\bf 70}(9), 1244-1247 (1993).
https://doi.org/10.1103/PhysRevLett.70.1244

\end{references}
\end{document}